# Smart Inventory Management System for Photovoltaic-Powered Freezer Using Wireless Sensor Network

**Janus Jade A. Basa[1], Patrick Lourenz G. Cu[2], Nathaniel N. Malabag[3], Luigi Angelo V. Naag[4], Dan Frederico P. Abacco[5], Mar Jun M. Siquihod[6], Gilfred Allen Madrigal[7], Lean Karlo S. Tolentino[8]**

[1,2,3,4,5,6,7,8] Department of Electronics Engineering, Technological University of the Philippines, Manila, Philippines

[8] University Extension Services Office, Technological University of the Philippines, Manila, Philippines

**ABSTRACT**

An inventory management system for the freezer powered by photovoltaic panels was developed in this study. It aims to promote energy efficiency and the responsible use of food. Its sensor network is an Arduino-based wireless network of sensors on a solar-powered freezer that is used to develop a smart inventory management system that is accessible and is easy to use. By having network of sensors implemented inside the freezer, the inventory of perishable and non-perishable items can easily be monitored without having to physically check the inside of the freezer. In connection with this, a complemental Android application was developed that would receive and display the data sent from the sensor network through GSM Shield SIM800L.

## 1. INTRODUCTION

Globally, about one third of human food produced is wasted which is roughly about 1.3 billion tons per year [1]. Food waste increases losses in financial and global natural resources. The depletion of this natural resources is not justified by the consumption efficiency of produced food from it. Food losses from those developing countries almost do not differ from those industrialized countries, the food losses that occur in post-harvest and processing levels is more than 40%, whereas in industrialized countries, about 40% of food losses takes place in retail and consumer levels [2]. The introduction of intelligent way of retail food management is necessary to mitigate the problem of food waste.

In a recent survey conducted in Italy [3] that puts light to food waste unveiled that 48.2% of food waste past due the expiration date, whereas 36.7% and 11.5% had been left forgotten in the fridge or pantry for too long. A possible solution has come up from the same survey to mitigate food waste. It showed 46.5% said that intelligent refrigerators would be helpful in planning a shopping list, automatically determining what is in stock or not and what is approaching end of shelf life. Respondents' suggestions were identified: it would be helpful to receive recipes for leftovers; they like to get advisories on how to best preserve and handle their food; and several would like to receive information on the shelf life and freshness of its contents. Most respondents stated they would prefer to use modern technology to receive this information, mainly by e-mail and software application for smartphone.

By the socio-economic problems stated above and taking advantage of the survey results, this project aims to turn a traditional chest freezer into a smart freezer, achieving a good food management and storage, eco-efficiency and efficient energy efficiency. The features implemented is not only to mitigate food waste but also to utilize solar energy by using photovoltaic panel. They are designed not only for domestic use but also for commercial and catering industries, as the system is easy to reproduce in a large scale. The smart food management and energy efficiency is the primary core of this system.

## 2. REVIEW OF RELATED LITERATURE

The smart refrigerator in [4] is a standard refrigerator that is fitted with a host of sensors and an image processing software for inventory and item recognition is an improved vision-based object detection and recognition of food inventory for the previous smart fridge. One drawback of its system is it cannot detect multiple items and object recognition was done by template matching which causes the image processing application to crash.

Meanwhile, [5] used a sensor network system that can monitor the stocks inside the refrigerator wirelessly. All data and images will be processed to provide the user an Internet of Things (IoT) [6]-[10] application through the cloud-based website Temboo. Temboo will have access to send data to the Dropbox. The user can monitor the stocks of the fridge wirelessly using Android Application.

The HighChest [11] is an innovative smart freezer designed to promote energy efficient behavior and the responsible storage of food. It features smart services such as weight sensing, temperature sensing, barcode reading matched with manual recording of expiration dates of its contents all accessible through a tablet attached externally to the chest freezer.

## 3. METHODOLOGY

The core of the study is sensor networks which are used to detect the amount of the contents and the internal temperature of the chest freezer. It uses cellular technology by GSM module connected to an Arduino to access required inventory information by the user.

### 3.1 Experimental Setup

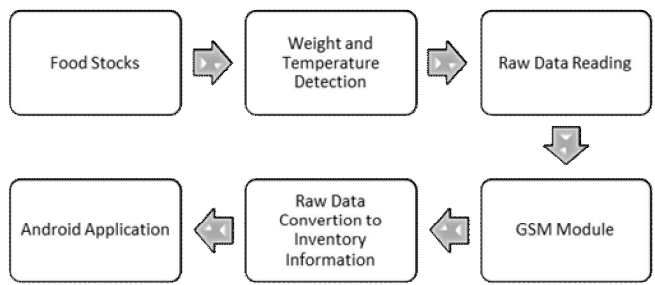

**Figure 1:** Process Flow of the System

The block diagram above (figure 3) shows how the system works. The food stock as the input to the system as weight. The weight of the food stock is measured using loadcell and the temperature is determined by using a DS18B20 digital temperature sensor. The sensor network continuously reads detect the weight and temperature, upon request by the user through a phone call the GSM module transmits the real-time raw data from the sensor network to smartphone of the user by SMS, the android application then converts the raw data into an information comprehensible to the user – the number of food stock inside the chest freezer.

### 3.2 Hardware Design

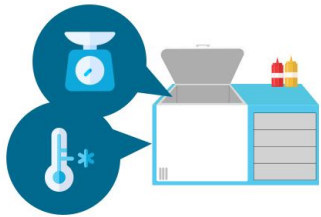

**Figure 2:** Overview of the Hardware Setup

The figure 2 above shows the setup of the chest freezer and sensor network inside. The system is composed of Arduino microcontroller attached to it is a GSM module, digital temperature sensor DS18B20, a straight bar load cell with maximum capacity of 20Kg for the elevated part of the chest freezer, four micro loadcells with a maximum load capacity of 200kg, 2 HX711 load cell amplifier breakout board. The platform of the loadcell is made of acrylic glass.

### 3.3 Circuit Diagram

Figure 3 below is the actual wiring of four micro load cell connected as a Wheatstone bridge to load cell amplifier board to an Arduino with a maximum capacity of 200kg. The platform is made of acrylic glass or plexiglass.

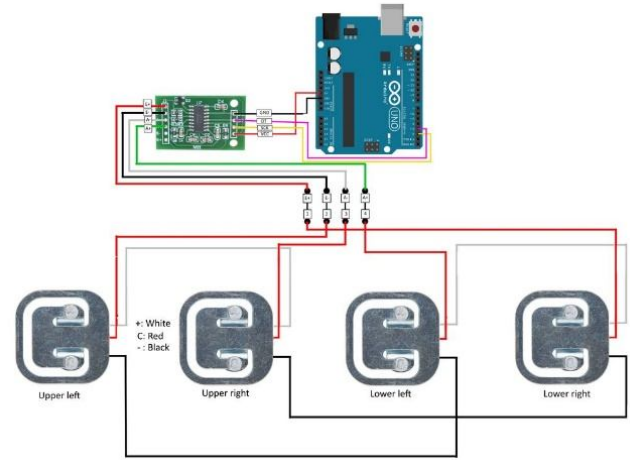

**Figure 3:** Actual Wiring for Load Cell 200 kg

Figure 4 below is the actual wiring of straight bar load cell connected to a load cell amplifier board to an Arduino with a maximum capacity of 20kg. The platform is made of acrylic glass or plexiglass.

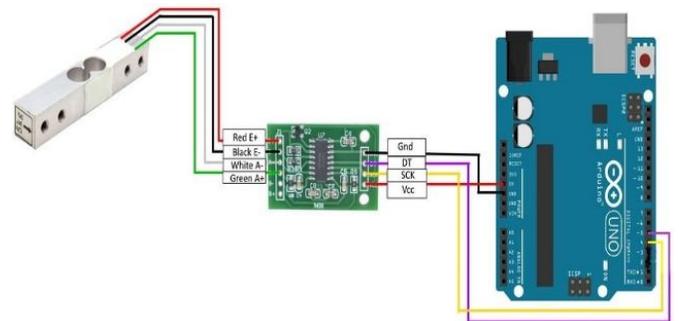

**Figure 4:** Actual Wiring for Load Cell 20 kg

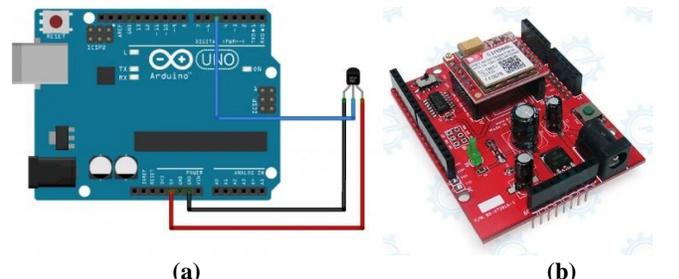

(a)          (b)
**Figure 5:** (a) Actual Wiring of DS18B20 Digital Temperature. (b) GSM Shield SIM800L

The DS18B20 connected to the Arduino this produces, this temperature sensor produces digital signal and a pull-up resistor is not required to be connected to data pin, the Arduino is already configured mimic a pull-up resistor connected from VCC pin to Data pin. The GSM Shield SIM800L is the communication link between the sensor network and the Android application.

### 3.4 Flow Chart

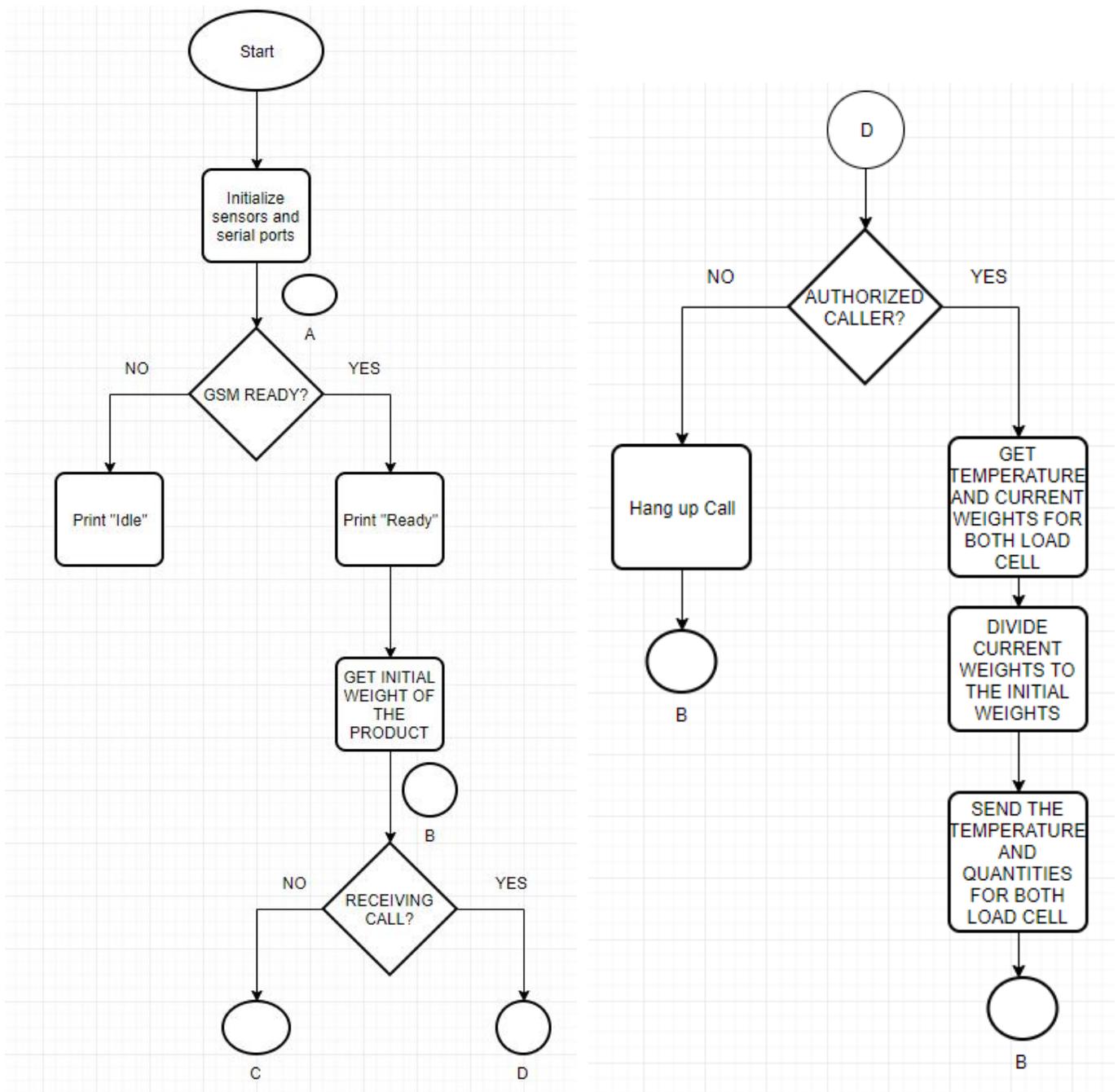

**Figure 6:** Flow Chart of Wireless Sensor Network to Android Application

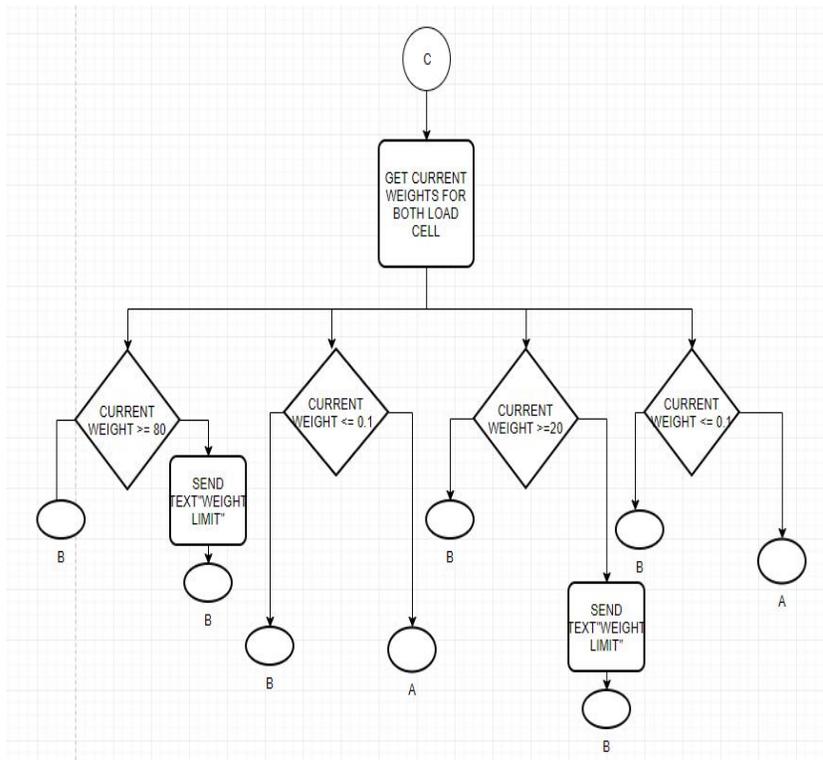

**Figure 7:** Flow Chart of Wireless Sensor Network to Android Application, continued.

Figures 6 and 7 shows the flow chart of the wireless sensor network to the Android application. The SIM800L GSM shield module will act as the communication link between the freezer and the Android app. As an authorized user clicks the "Check" button, it will give a call to the SIM800L GSM Shield module which will prompt a process that will get the quantity of the stocks inside the freezer as well as the temperature. The Arduino will timely monitor the weight of the products to check whether it exceeded the limits which are 20kg for the elevated platform and 80kg for the other platform. Once either the two (2) exceeded the limit, a text message (SMS) coming from the GSM module will be sent notifying the owner that they had exceeded the limit.

## 4. RESULTS AND DISCUSSIONS

The Smart Inventory Management System was tested to assess the accuracy of the temperature and weight sensors and the responsiveness of the GSM shield SIM800L. The sensors` accuracies greatly affect the performance of the system since the weight should be check occasionally for the freezer has a weight capacity of 100Kg as well as the maintaining temperature of the stored food thus temperature is significant.

The first load cell (100kg) was calibrated with a calibration factor of -9475 and an offset scale of -922696 using a sample weight of 1kg, 10kg, 20kg, 30kg and 40kg. In Table 1, it shows that the percentage error is 0, .601, .451, 0 and .049 for weights 1kg, 10kg, 20kg, 30kg, and 40kg respectively in which the success rate in measuring the exact weight is approximately 99.5%.

**Table 1:** Percentage errors accounted for 5 samples for load cell (200 kg)

| Actual Weight | Measured Weight | Percent Error |
|---|---|---|
| 1 kg | 1 kg | 0 |
| 10 kg | 9.94 kg | 0.602 |
| 20 kg | 19.91 kg | 0.451 |
| 30 kg | 30 kg | 0 |
| 40 kg | 40.02 kg | 0.05 |

The second load cell (20kg) was calibrated with a calibration factor of 40100 and an offset scale of 190468 using a sample weight of 1 kg, 5 kg, 10k g, 15 kg and 20 kg. In table 2, it shows that the percentage error is 1.98019802, 1.816347124, .0, 0.066644452 and .02002002 for weights 1 kg, 5 kg, 10 kg, 15 kg, and 20 kg respectively in which the success rate in measuring the exact weight is approximately 99.2%.

**Table 2:** Percentage errors accounted for 5 samples for load cell (20 kg)

| Actual Weight | Measured Weight | Percent Error |
|---|---|---|
| 1 kg | 1.02 kg | 1.98 |
| 5 kg | 4.91 kg | 1.816 |
| 10 kg | 10 kg | 0 |
| 15 kg | 15.01 kg | 0.067 |
| 20 kg | 19.96 kg | 0.2 |

The temperature sensor was calibrated in five different temperatures of 36 ⁰, 20 ⁰, 10 ⁰, 0 ⁰, and -10 ⁰. In table 3, it shows that the percentage error is 0, 5.128, 11.321, 0, and 10.526 for 36 ⁰, 20 ⁰, 10 ⁰, 0 ⁰, and -10 ⁰ respectively in which the success of rate in measuring the exact temperature is approximately 95%.

**Table 3:** Percentage errors accounted for 5 samples for DS18B20

| Actual Temperature | Measured Temperature | Percent Error |
|---|---|---|
| 36 °C | 36 °C | 0 |
| 20 °C | 18.974 °C | 5.128 |
| 10 °C | 8.868 °C | 11.321 |
| 0 °C | 0 °C | 0 |
| -10 °C | -8.947 °C | 10.526 |

## 5. CONCLUSION

With the increase of adoption of solar-powered freezer into the home environment especially on remote areas, the presence of solution for food and beverages management is essential for inventory and monitoring goods inside the freezer, even more so than in refrigerators. In this paper, the Smart Inventory Management System for Photovoltaic-Powered Freezer Using Wireless Sensor Network is presented with the purpose of monitoring the temperature and quantity of the goods inside the freezer.

The system is very efficient since it is dependent on several factors which are the accuracy of the sensors and the responsiveness of the GSM shield SIM800L. For the accuracy of the sensors, the temperature and weight sensors are calibrated on certified temperature and weights and showed a negligible percentage error. In addition, the responsiveness of GSM shield SIM800L is based on how strong the transmission and reception of text messages and calls are in your area.

An image processing software for inventory and item recognition could be designed as a future activity to enhance the capability of the system to handle several types of items. Likewise, other methods for the communication link between the sensor network and Android application could be implemented for faster acquisition of data. The wireless sensor network for Smart Inventory Management system could contribute for easier handling and obtaining precise quantity of the stored item.

## REFERENCES

1. S. Caronna. **Report on How to Avoid Food Wastage: Strategies for Improving the Efficiency of the Food Chain in the EU**. Committee on Agriculture and Rural Development, European Parliament; Brussels, Belgium. 2011
2. FAO. **Global Food Losses and Food Waste—Extent, Causes and Prevention**. FAO; Rome, Italy: 2011
3. S. Gaiani, **Lo spreco alimentare domestico in Italia: stime, cause ed impatti,** PhD diss., alma, 2013.
4. J. K. P. Aranilla, T. A. C. Dela Fuente, T. Y. Quintos, E. O. Samonte, J. P. Ilao, and F. P. Lai. **Liveitup! 2 Smart Refrigerator: Improving Inventory Identification and Recognition,** in *Research Congress 2013 De La Salle University Manila,* pp. 1-9, 2013.
5. J. Velasco, L. Alberto, H. D. Ambatali, M. Canilang, V. Daria, J. B. Liwanag, G. A. Madrigal, E. Galido, and L. K. Tolentino. **Internet of Things-based (IoT) Inventory Monitoring Refrigerator using Arduino Sensor Network,** *Indonesian Journal of Electrical Engineering and Computer Science,* In press.
6. B. N. Fortaleza, R. O. Serfa Juan, and L. K. S. Tolentino. **IoT-based Pico-Hydro Power Generation System Using Pelton Turbine**, *Journal of Telecommunication, Electronic and Computer Engineering (JTEC)*, vol. 10, no. 1-4, pp. 189-192, 2018.
7. R. T. M. Cruz, L. K. S. Tolentino, R. O. Serfa Juan, and H. S. Kim. **IoT-based monitoring model for pre-cognitive impairment using pH level as analyte**, *International Journal of Engineering Research and Technology*, vol. 12, no. 5, pp. 711-718, 2019.
8. L. K. Tolentino, et al. **AQUADROID: AN APP FOR AQUAPONICS CONTROL AND MONITORING**, in *6th International Conference on Civil Engineering (6th ICCE 2017)*, pp. 1-8, 2017.
9. A D. M. Africa, G. Ching, K. Go, R. Evidente, and J. Uy. **A Comprehensive Study on Application Development Software Systems,** *International Journal of Emerging Trends in Engineering Research,* vol. 7, no. 8, pp. 99-103, 2019.
https://doi.org/10.30534/ijeter/2019/03782019
10. E. R. Magsino. **Energy Monitoring System Incorporating Energy Profiling and Predictive Household Movement for Energy Anomaly Detection,** *International Journal of Emerging Trends in Engineering Research,* vol. 7, no. 8, pp. 151-156, 2019.
https://doi.org/10.30534/ijeter/2019/08782019
11. M. Bonaccorsi, S. Betti, G. Rateni, D. Esposito, A. Brischetto, M. Marseglia, P. Dario, and F. Cavallo. **'HighChest': An Augmented Freezer Designed for Smart Food Management and Promotion of Eco-Efficient Behaviour**, *Sensors*, vol. 17, no. 6, pp. 1-21, 2017.
https://doi.org/10.3390/s17061357